\documentclass[10pt,final,conference]{IEEEtran}
\usepackage[latin9]{inputenc}
\usepackage{units}
\usepackage{amsthm}
\usepackage{amsmath}
\usepackage{amssymb}
\usepackage{esint}
\usepackage[unicode=true,pdfusetitle,
 bookmarks=true,bookmarksnumbered=false,bookmarksopen=false,
 breaklinks=false,pdfborder={0 0 1},backref=page,colorlinks=false]
 {hyperref}

\makeatletter
\theoremstyle{plain}
\newtheorem{thm}{\protect\theoremname}
\theoremstyle{definition}
\newtheorem{example}[thm]{\protect\examplename}
\theoremstyle{plain}
\newtheorem{cor}[thm]{\protect\corollaryname}
\theoremstyle{plain}
\newtheorem{lem}[thm]{\protect\lemmaname}

\usepackage{amsfonts}

\newcommand{\commentout}[1]{}

\makeatother

\providecommand{\corollaryname}{Corollary}
\providecommand{\examplename}{Example}
\providecommand{\lemmaname}{Lemma}
\providecommand{\theoremname}{Theorem}

\begin{document}

\title{Extendable MDL}

\author{Peter Harremo{\"e}s\linebreak{}
{\small Copenhagen Business College}\linebreak{}
{\small Copenhagen, Denmark}\linebreak{}
{\small E-mail: harremoes@ieee.org}}
\maketitle
\begin{abstract}
In this paper we show that a combination of the minimum description
length principle and an exchange-ability condition leads directly
to the use of Jeffreys prior. This approach works in most cases even
when Jeffreys prior cannot be normalized. Kraft's inequality links
codes and distributions but a closer look at this inequality demonstrates
that this link only makes sense when sequences are considered as prefixes
of potential longer sequences. For technical reasons only results
for exponential families are stated. Results on when Jeffreys prior
can be normalized after conditioning on a initializing string are
given. An exotic case where no initial string allow Jeffreys prior
to be normalized is given and some way of handling such exotic cases
are discussed.
\end{abstract}

\section{Introduction}

A major problem in Bayesian statistics is to assign prior distributions
and to justify the choice of prior. The minimum description length
(MDL) approach to statistics is often able to overcome this problem,
but although MDL may look quite similar to Bayesian statistics the
inference is different. One of the main results in MDL is that Jeffreys
prior is asymptotically minimax optimal with respect to both redundancy
and regret. Despite this positive result there are two serious technical
complications that we will address in this paper.

The first complication is that in MDL the use of a code based on Jeffreys
prior is normally considered as suboptimal to the use of the normalized
maximum likelihood distribution (NML). Jeffreys prior turn out to
be optimal if we make a more sequential approach to online prediction
and coding. The key idea is to consider \emph{extended sequences}.

The second complication is that in many important applications, Jeffreys
prior cannot be normalized. When Jeffreys prior cannot be normalized
it is often (but not always) the case that the Shtarkov integral is
infinite so that the NML distribution does not exist. This problem
is often handled by conditioning by a short sequence of initial data.
In Bayesian statistics this has lead to a widespread use of \emph{improper
prior distributions} and in MDL it has lead to the definition of the
SNML predictor. Our sequential approach will justify the use of improper
Jeffreys priors and describe in which sense the use of improper Jeffreys
distributions is normally preferable to the SNML predictor.

In the classical frequential approach to statistics a finite sequence
is considered as a sub-sequence of an infinite sequence. I Bayesian
statistics a finite sequence is normally considered without reference
to longer sequences. In this paper we will take a standpoint in between.
We will think of a finite sequence as a prefix of \emph{potentially
longer finite sequences}. Only in this way we can justify the equivalence
between codes and distributions via Kraft's inequality. In this short
paper we shall restrict our attention to exponential families to avoid
technical complications related to measurablity etc. Despite this
restriction our results cover many important applications and the
model is still sufficiently flexible to illustrate ideas that can
be generalized to a more abstract setting.

The rest of this paper is organized as follows. In Section \ref{sec:Preliminaries}
notation is fixed and some well-known basic results are stated in
the way that we are going to use them. In Section \ref{sec:MDL-and-Kraft's}
we will see that the use of Kraft's inequality is relevant if we consider
short sequences as sub-sequences of longer sequences. In Section \ref{sec:Improper-priors}
we define exponential prediction systems and we will see how such
systems are given by prior measures on the parameter space and for
which sequences conditional distributions exists. In Section \ref{sec:Jeffreys-prior}
the optimality of Jeffreys prior is described and some results on
when conditional distributions exists are stated. These sections are
given in the logical order of reasoning but they can be read quite
independently. In the proceeding version of this paper most proofs
have been left out. A longer version of this paper with an appendix
that contains proofs of all theorems, can be found on arXiv.org .
The paper ends with a short discussion.

\section{Preliminaries\label{sec:Preliminaries}}

\subsection{Definitions for exponential families}

The exponential family $\{P_{\beta}\mid\beta\in\Gamma^{\text{can}}\}$
based on the probability measure $P_{0}$ is given in a canonical
parametrization, 
\begin{equation}
\frac{\mathrm{d}P_{\beta}}{\mathrm{d}P_{0}}=\frac{\exp\left(\beta x\right)}{Z(\beta)},\,\beta\in\Gamma^{\text{can}}\label{eq:expdef}
\end{equation}
where $Z$ is the partition function $Z(\beta)=\int\exp(\beta x)\,\mathrm{d}P_{0}x$,
and $\Gamma^{\text{can}}:=\{\beta\mid Z(\beta)<\infty\}$ is the \emph{canonical
parameter space}. Note that we allow the measure $P_{0}$ to have
both discrete and continuous components. The trivial case where $\Gamma^{\text{can}}$
has no interior points is excluded from the analysis. In Equation
\ref{eq:expdef} $\beta x$ will denote the product of real numbers
when the exponential family is 1-dimensional and $\beta x$ will denote
a scalar product when the exponential family has dimension $d>1$
so that $\beta$ and $x$ are vectors in $\mathbb{R}^{d}$. See \cite{Barndorff-Nielsen1978}
for more details on exponential families.

For our problem it is natural to work with \emph{extended exponential
families} as defined in \cite{Csiszar2003}. For a probability distribution
$Q$ on $\mathbb{R}^{k}$ the convex support $cs\left(Q\right)$ is
the intersection of all convex \emph{closed} sets that have $Q$-probability
1. The convex core $cc\left(Q\right)$ is the intersection of all
convex \emph{measurable} sets with $Q$-probability 1, \cite{Csiszar2001}.
We have $cc\left(Q\right)\subseteq cs\left(Q\right).$ An extreme
point $x$ in $cs\left(Q\right)$ belongs to $cc\left(Q\right)$ if
and only if $Q\left(x\right)>0.$ In its mean value parametrization
the exponential family based on a measure with bounded support has
a natural extension to $cc\left(Q\right).$ In particular $\delta_{x}$
belongs to the extended exponential family if $Q$ has a point mass
in $x$ and $x$ is an extreme point of $cs\left(Q\right).$

The elements of the exponential family are also parametrized by their
mean value $\mu$. We write $\mu_{\beta}$ for the mean value corresponding
to the canonical parameter $\beta$ and $\hat{\beta}\left(\mu\right)$
for the canonical parameter corresponding to the mean value $\mu.$
Note that we allow infinite values of the mean. The element in the
exponential family with mean $\mu$ is denoted $P^{\mu}.$ The mean
value range $M$ of the exponential family is the range of $\beta\rightarrow\mu_{\beta}$
and is a subset of the convex core. For 1-dimensional families we
write $\mu_{\sup}=\sup M$, and $\mu_{\inf}=\inf M$. If $P_{0}$
has a point mass at $\mu_{\inf}>-\infty$ and the support of $P_{0}$
is a subset of $\left[\mu_{\inf},\infty\right[,$ then the exponential
family is extended by the element $P_{-\infty}=P^{\mu_{\inf}}=\delta_{\mu_{\inf}}$,
and likewise the exponential family is extended if $Q$ has a point
mass in $\mu_{\sup}<\infty$ and the support of $Q$ is a subset of
$\left]\textrm{-}\infty,\mu_{\sup}\right].$ For any $x$ the distribution
$P_{\hat{\beta}\left(x\right)}=P^{x}$ is the maximum likelihood distribution. 

The covariance function $V$ is the function that maps $\mu\in M$
into the covariance of $P^{\mu}.$ If $M$ has interior points then
the exponential family is uniquely determined by its covariance function.
The Fisher information of an exponential family in its canonical parametrization
is $I_{\beta}=V\left(\mu_{\beta}\right)$ and the Fisher information
of the exponential family in its mean value parametrization is $I^{\mu}=V\left(\mu\right)^{\textrm{-}1}.$ 

For elements of an exponential family we introduce \emph{information
divergence} as
\begin{align*}
D\left(\left.x\right\Vert y\right): & =D\left(\left.P^{x}\right\Vert P^{y}\right)=\int\ln\left(\frac{\mathrm{d}P^{x}}{\mathrm{d}P^{y}}\right)\,\mathrm{d}P^{x}.
\end{align*}
This defines a \emph{Bregman divergence} on the mean value range and
under some regularity conditions this Bregman divergence uniquely
characterizes the exponential family \cite{Banerjee2005}.

\subsection{Posterior distributions}

If the mean value parameter has prior distribution $\nu$ and $x$
has been observed then the posterior distribution has density 
\[
\frac{\mathrm{d}\nu\left(\cdot\vert x\right)}{\mathrm{d}\nu}\left(y\right)\sim\exp\left(-D\left(\left.x\right\Vert y\right)\right).
\]

\textbf{Notation} We use $x^{m}$ to denote $\left(x_{1},x_{2},\dots,x_{m}\right)$
and $x_{m}^{n}$ to denote $\left(x_{m},x_{m+1},\dots,x_{n}\right).$
We use $\tau$ as short for $2\pi$ and $\sim$ to denote that two
functions or measures are proportional.

If a sequence $x_{1},x_{2},\dots,x_{m}$ has been observed then the
posterior distribution has density 
\begin{multline*}
\frac{\mathrm{d}\nu\left(\cdot\vert x^{m}\right)}{\mathrm{d}\nu}\left(y\right)\sim\prod_{i=1}^{m}\exp\left(\textrm{-}D\left(\left.x_{i}\right\Vert y\right)\right)\\
=\left(\prod_{i=1}^{m}\exp\left(-D\left(\left.x_{i}\right\Vert \bar{x}\right)\right)\right)\cdot\exp\left(\textrm{-}nD\left(\left.\bar{x}\right\Vert y\right)\right)
\end{multline*}
where $\bar{x}$ denotes the average of the sequence $x^{m}$, where
we have an equality that is of general validity for Bregman divergences.
Since the first factor does not depend on $y$ we have 
\[
\frac{d\nu\left(\cdot\left|x^{m}\right.\right)}{d\nu}\left(y\right)\sim\exp\left(-mD\left(\left.\bar{x}\right\Vert y\right)\right).
\]

\subsection{MDL in exponential families}

For some exponential families the \emph{minimax regret} $C_{\infty}$
is finite. See \cite{Grunwald2007} for details about how this quantity
is defined. If $C_{\infty}$ is finite the minimax regret is assumed
if we code according to the \emph{NML distribution}. In general the
optimal code for $X_{1}$ will depend on whether the sample size is
$n=1$ or whether $X_{1}$ is considered as a sub-sequence of $X^{n}.$
In cases where $C_{\infty}$ is infinite one may use a conditional
versions instead such as \emph{sequential NML} (SNML).

Of central importance for our approach are result developed by Barron,
Rissanen et al. that if the parameter space of an exponential family
is restricted to a non-empty compact subset of the interior of the
convex core, then the minimax regret is finite and equal to 
\begin{equation}
C_{\infty}=\frac{d}{2}\ln\frac{n}{\tau}+\ln J+o(1),\label{eq:asymp}
\end{equation}
where $J$ denotes the \emph{Jeffreys integral}
\begin{equation}
J=\int_{\Gamma^{\text{can}}}\left(\det I_{\beta}\right)^{\nicefrac{1}{2}}\,\mathrm{d}\beta=\int_{M}\left(\det V\left(x\right)\right)^{\textrm{-}\nicefrac{1}{2}}\,\mathrm{d}x.~\label{eq:jeffreysintegral}
\end{equation}
where $I_{\beta}$ denotes the \emph{Fisher information matrix} \cite{Grunwald2007}.
Moreover, the same asymptotic regret (\ref{eq:asymp}) is achieved
by the Bayesian marginal distribution equipped with Jeffreys prior.
In MDL this result is often used as the most important reason for
using \emph{Jeffreys prior} with density $w(\mu)=\left(\det V\left(\mu\right)\right)^{\textrm{-}\nicefrac{1}{2}}/J$.
The use of the NML predictor requires knowledge of the sample size
and the performance of the SNML predictor will depend on the order
of the observations except if it corresponds to the use of Jeffreys
prior \cite{Hedayati2012}. 

If the parameter space is restricted to a non-empty compact subset
of the interior of the convex core (called an \emph{ineccsi} set in
\cite{Grunwald2007}) the Jefftreys integral is automatically finite
but typically there is no natural way of restricting the parameter
space in applications and in most cases the Jeffreys integral is infinite.
It thus becomes quite relevant to investigate what happens if the
parameter space is \emph{not} restricted to an ineccsi set. To answer
this question, one needs to know when the Jeffreys integral is finite,
and how to handle situations where Jeffreys integral is not finite.

\subsection{Exchangeability, sufficiency, and consistency}

Prediction in exponential families satisfy the exchangability condition
that the probability of sequence does not depend on the order of the
elements. We may also say the predictor is invariant under permutations
of the elements in a sequence. The importance of this exchangablity
condition in MDL was emphasized in \cite{Hedayati2012}. A related
but more important type of exchangablity is that the probability of
a sequence given a sub sequence $x^{j}$ does not depend on the order
of the observations in the sub-sequence. A stronger requirement is
that the predicted probability of a sequence given a sub-sequence
$x^{j}$ only depends the average $\bar{x}$ of the subsequence, i.e.
the sample average is a \emph{sufficient statistic}. We are also interested
in consistency of the system of predictors. Note that $P\left(\left.x_{\ell+1}^{n}\right|x^{\ell}\right)=P\left(\left.x^{n}\right|x^{\ell}\right).$
A system of predictors is \emph{consistent} if the prediction $P\left(\left.x^{n}\right|x^{\ell}\right)=P\left(\left.x^{n}\right|x^{m}\right)\cdot P\left(\left.x^{m}\right|x^{\ell}\right).$
A consistent system of predictors is generated from predictions of
the next symbol given by the past symbols.

\section{MDL and Kraft's inequality\label{sec:MDL-and-Kraft's}}

We recall that a code is uniquely decodable if any finite sequence
of input symbols give a unique sequence of output symbols. It is well-known
that a uniquely decodable code satisfies Kraft's inequality 
\begin{equation}
\sum_{a\in\mathbb{A}}\beta^{\textrm{-}\ell\left(a\right)}\leq1\label{eq:Kraft}
\end{equation}
where $\ell\left(a\right)$ denotes the length of the codeword corresponding
to the input symbol $a\in\mathbb{A}$ and $\beta$ denotes the size
of the output alphabet. The length of a codeword is an integer. Normally
the use of non-integer valued code length functions is justified by
reference to the noiseless coding theorem which require some interpretation
of the notion of probability distributions and their mean values.
To emphasize our sequential point of view we formulate a version of
Kraft's inequality that allow the code length function to be non-integer
valued.
\begin{thm}
\label{thm:Kraft}Let $\ell:\mathbb{A}\rightarrow\mathbb{R}$ be a
function. Then the function $\ell$ satisfies Kraft's inequality (\ref{eq:Kraft})
if and only if for all $\varepsilon>0$ there exists an integer $n$
and a uniquely decodable fixed-to-variable length block code $\kappa:\mathbb{A}^{n}\rightarrow\mathbb{B}^{\ast}$
such that
\[
\left\vert \bar{\ell}_{\kappa}\left(a^{n}\right)-\frac{1}{n}\sum_{i=1}^{n}\ell\left(a_{i}\right)\right\vert \leq\varepsilon
\]
where $\bar{\ell}_{\kappa}\left(a^{n}\right)$ denotes the length
$\ell_{\kappa}\left(a^{n}\right)$ divided by $n.$ The uniquely decodable
block code can be chosen to be prefix free.
\end{thm}
It is only possible to obtain a unique correspondence between code
length functions and (discrete) probability measures by considering
codewords as prefixes of potentially longer codewords. If we restrict
our attention to code words of some finite fixed length then Kraft's
inequality does not give a necessary and sufficient condition of decodability.
Like in Bayesian statistics we focus on finite sequences. Contrary
to Bayesian statistics we should always consider a finite sequence
as a prefix of \emph{longer finite} sequences. Contrary to frequential
statistics we do not have to consider a finite sequence as a prefix
of an \emph{infinite} sequence.

If the set of input symbols is not discrete one has to introduce some
type of distortion measure, but we will abstain from discussing this
complication in this short note.

\section{Improper priors\label{sec:Improper-priors}}

In this section we will talk about a prior measure even when it cannot
be normalized and we will call it a \emph{proper prior} when it can
be normalized to a probability measure.

\subsection{Finiteness structure}

If a sequence of length $m$ with average $\bar{x}$ is observed then
the prior integral is either finite or infinite. Let $F_{m}$ denote
the subset of average values in the convex core such that the prior
integral is finite for samples of size $m.$
\begin{thm}
\label{thm:The-sets-convex}The sets $F_{m}$ form an increasing sequence
of convex subsets of the convex core, i.e. $F_{1}\subseteq F_{2}\subseteq F_{3}\subseteq\dots cc$.\end{thm}
\begin{example}
Consider the Gaussian location family. For this family $D\left(y\Vert x\right)=\frac{\left(x-y\right)^{2}}{2}.$
If the prior has density $\exp\left(\alpha x^{2}\right),$ then the
prior can be normalized to a posterior distribution when
\[
\intop_{-\infty}^{\infty}\exp\left(\alpha x^{2}\right)\exp\left(\textrm{-}m\frac{\left(x-y\right)^{2}}{2}\right)\,\mathrm{d}x
\]
so the integral is finite when $m>2\alpha.$ 

If the prior has density $\exp\left(x^{4}\right)$ then there exists
no $m$ for which the prior can be normalized. \end{example}
\begin{thm}
\label{thm:Conv}Assume that $x_{1}\in F_{m}$ and $\mu_{0}$ is in
the convex core. Then $\left(1-\frac{m}{n}\right)x_{0}+\frac{m}{n}\mu_{1}\in F_{n}$.
\end{thm}
An important special case is when the convex core equals $\mathbb{R}^{d}$.
In this case we have that if $F_{n}\neq\emptyset$ then $F_{n+1}=\mathbb{R}^{k}$.

The next example shows that Theorem \ref{thm:Conv} is 'tight'.
\begin{example}
The family of exponential distributions has $D\left(\lambda\Vert\mu\right)=\frac{\lambda}{\mu}-1-\ln\frac{\lambda}{\mu}.$
Consider the prior density $\exp\left(x^{\textrm{-}1}\right)\cdot x^{\textrm{-}2}.$
The conditional integral is 
\[
\int_{0}^{\infty}\exp\left(x^{\textrm{-}1}\right)x^{\textrm{-}2}\cdot\exp\left(\textrm{-}m\left(\frac{\bar{x}}{x}-1-\ln\frac{\bar{x}}{x}\right)\right)\,\mathrm{d}x.
\]
The integral $\int_{1}^{\infty}\exp\left(x^{\textrm{-}1}\right)\cdot x^{\textrm{-}2}\,\mathrm{d}x$
is finite so we only have to consider the integral
\begin{multline*}
\int_{0}^{1}\exp\left(x^{\textrm{-}1}\right)x^{\textrm{-}2}\cdot\exp\left(\textrm{-}m\left(\frac{\bar{x}}{x}-1-\ln\frac{\bar{x}}{x}\right)\right)\,\mathrm{d}x\\
=\bar{x}^{n}\exp\left(n\right)\int_{0}^{1}\exp\left(\left(1-m\bar{x}\right)x^{\textrm{-}1}\right)\cdot x^{n}x^{\textrm{-}2}\,\mathrm{d}x\,.
\end{multline*}
The substitution $y=x^{\textrm{-}1}$ gives 
\begin{multline*}
\int_{0}^{1}\exp\left(\left(1-m\bar{x}\right)x^{\textrm{-}1}\right)\cdot x^{n}x^{\textrm{-}2}\,\mathrm{d}x\\
=\int_{1}^{\infty}\exp\left(\left(1-m\bar{x}\right)y\right)\cdot y^{\textrm{-}n}\,\mathrm{d}y\,.
\end{multline*}
We see that for $n>1$ the integral is finite if and only if $\bar{x}\geq\nicefrac{1}{m}$,
which implies that $F_{m}=\left[\nicefrac{1}{m},\infty\right[.$
\end{example}

\subsection{Existence of a prior}

We will now define an \emph{exponential prediction system}. We consider
a sequence of variables $X_{1},X_{2},\dots$with values in $\mathbb{R}^{d}$.
For some sequences of outcomes $x^{m}$ a probability measure $P\left(\cdot\left|x^{m}\right.\right)$
on $\mathbb{R}^{d}$ is given and the interpretation of this probability
measure is that it gives the probability or prediction of the next
variable $X_{m+1}$ given the values of the previous variables. Equivalently
we may think of $P\left(\cdot\left|x^{m}\right.\right)$ as an instruction
about how the next variable should be coded given the values of the
previous variables. Further we will assume that if $P\left(\cdot\vert x^{m}\right)$
is defined then $P\left(\cdot\vert x^{n}\right)$ is also defined
for any sequence $x^{n}$ with $x^{m}$ as prefix. Further we will
assume that the sum is sufficient for prediction, i.e. $P\left(\cdot\left|x^{m}\right.\right)$
only depends on the value of the sum $x_{1}+x_{2}+\dots+x_{m}$. 

An exponential prediction system as described above can be extended
to a consistent prediction system for sequences and we note that the
sum is still sufficient for predicting sequences.  Conversely, a consistent
prediction system for sequences can be reconstructed from its restriction
to predictions of the next symbol.

Assume that $P\left(\cdot\left|x^{m}\right.\right)$ exists. Then
we have a consistent system of probability measures on the variables
$X_{m+1},X_{m+2},\dots$ for which the sums of the previous variables
are sufficient statistics for the following variables. According to
results of S. Lauritzen any such system is a mixture of elements in
an exponential family when the predictor is defined even for initial
sequences of length $m=0$ \cite{Lauritzen1982a}. Therefore there
exists a measure $P_{0}$ and a probability measure $\nu_{x^{n}}$
over the convex core such that
\[
\frac{\mathrm{d}P\left(\cdot\left|x^{m}\right.\right)}{\mathrm{d}P_{0}}\left(x\right)=\int_{cc}\frac{\exp\left(x\cdot\hat{\beta}\left(y\right)\right)}{Z\left(\hat{\beta}\left(y\right)\right)}\,\mathrm{d}\nu_{x^{m}}y.
\]
These 'prior distributions' $\nu_{x^{n}}$ are updated to 'posterior
distributions' in the usual fashion
\[
\frac{\mathrm{d}\nu_{x^{m+1}}}{\mathrm{d}\nu_{x^{m}}}\left(x\right)\sim\exp\left(\textrm{-}D\left(\left.x_{m+1}\right\Vert x\right)\right).
\]
The following theorem extends results of S. Lauritzen to cases where
$m>0.$
\begin{thm}
\label{thm:Laur}For an exponential prediction system there exists
an exponential family based on a probability measure $P_{0}$ and
a prior measure $\eta$ over the mean value range $M$ of the exponential
family such that \textup{
\begin{multline*}
\frac{\mathrm{d}P\left(\cdot\left|x^{m}\right.\right)}{\mathrm{d}P_{0}}\left(x\right)=\\
\int_{M}\frac{\exp\left(x\cdot\hat{\beta}\left(z\right)\right)}{Z\left(\hat{\beta}\left(z\right)\right)}\frac{\exp\left(\textrm{-}mD\left(\left.\bar{x}\right\Vert z\right)\right)}{\int_{M}\exp\left(\textrm{-}mD\left(\left.\bar{x}\right\Vert z\right)\right)\,\mathrm{d}\eta z}\,\mathrm{d}\eta z.
\end{multline*}
}
\end{thm}

\section{Jeffreys prior\label{sec:Jeffreys-prior}}

\subsection{Conditional regret}

We will use conditional regret to evaluate the quality of a predictor.
For a conditional setup Peter Gr{\"u}nwald has defined three different
notions of conditional regret \cite[subsection 11.4.2]{Grunwald2007}.
First we assume that the sample space is finite. We let $P^{t}$ denote
a distribution in the exponential family and we compare it with a
predictor $Q\left(\cdot\vert\cdot\right).$ If a sequence $x^{n}$
is observed then the optimal code based on an element in the exponential
family would provide codelength $-\ln P^{t}\left(x^{n}\right).$ In
order to code the same sequence using a predictor $Q\left(\cdot\vert\cdot\right)$
when the initial string $x^{m}$ has been observed, the code length
for the rest of the sequence is $\textrm{-}\ln Q\left(\left.x^{n}\right|x^{m}\right)$.
The \emph{regret-2} is defined as the difference 
\[
\textrm{-}\ln Q\left(\left.x^{n}\right|x^{m}\right)-\left(\textrm{-}\ln P^{t}\left(x^{n}\right)\right).
\]
 If the optimal distribution from the exponential family is used the
regret of the predictor with respect to the sequence is 
\begin{align*}
REG_{Q}\left(\left.x^{n}\right|x^{m}\right) & =-\ln Q\left(\left.x^{n}\right|x^{m}\right)-\left(-\ln P^{\bar{x}}\left(x^{n}\right)\right).\\
 & =\ln\frac{P^{t}\left(x^{n}\right)}{Q\left(\left.x^{n}\right|x^{m}\right)}.
\end{align*}
If the sample space is not finite then we replace probabilities with
densities with respect to a fixed measure $P_{0}$ in the exponential
family. 

Kraft's ineqality implies that one code based on a probability measure
cannot have shorter codewords than another code for all outcomes.
The following theorem states that something similar holds for consistent
predictors. 
\begin{thm}
\label{thm:relative}Let $Q_{1}$ and $Q_{2}$ denote two different
exponential prediction systems for the same exponential family. Then
there exist a sequence $x_{1},x_{2},\dots$ and a number $m$ such
that
\[
\lim_{n\to\infty}\inf\left(REG_{Q_{2}}\left(\left.x^{n}\right|x^{m}\right)-REG_{Q_{1}}\left(\left.x^{n}\right|x^{m}\right)\right)>0.
\]

\end{thm}

\subsection{Optimality of Jeffreys prior}

We are now able to combine our sequential approach with existing results
on optimality of Jeffreys prior. 
\begin{thm}
\label{thm:taet}Assume that $\left(P^{x}\right)$ is a exponential
family based on the probability measure $P_{0}$ and that $Q\left(\cdot\vert\cdot\right)$
denotes an exponential prediction system based on the probability
measure $Q_{0}$ with prior measure $\nu$ on the mean value range
$M$. 

If $Q_{0}=P_{0}$ and the support of the prior measure $\nu$ equals
the closure of the mean value range of the exponential family, then
for any $P^{x}$ in the extended exponential family with $x$ in the
convex core and any sequence $x_{1},x_{2},\dots$ satisfying 
\[
\lim\inf D\left(\left.P^{\bar{x}}\right\Vert P^{x}\right)>0
\]
then the conditional regret-2 of the exponential prediction system
$Q\left(\cdot\vert\cdot\right)$ is eventually less than the conditional
regret of $P^{x}$ with respect to the sequence $x_{1},x_{2},\dots$ 

Exponential prediction systems based on $P_{0}$ and with dense prior
are the only exponential prediction systems satisfying this property.
\end{thm}
Further conditions are needed in order to single out the Jeffreys
prior. The conditional Jeffreys integral is defined as
\[
J\mid x^{m}=\int\frac{\exp\left(\textrm{-}mD\left(\left.P^{\bar{x}}\right\Vert P^{x}\right)\right)}{\left(\det V\left(x\right)\right)^{\nicefrac{1}{2}}}\,\mathrm{d}x
\]
where $\bar{x}$ is the sample average of $x^{m}$. The following
theorem states that an exponential prediction system is asymptotically
optimal with respect to minimax regret if and only if it is based
on Jeffreys prior. A proof of essentially the same theorem can be
found in \cite{Grunwald2007}.
\begin{thm}
\label{thm:Jeffreys} If an exponential prediction system $Q$ is
based on Jeffreys prior and an element $P^{x}$ in the exponential
family corresponding to an interior point $x$ in the convex core
and $x_{1}x_{2}\dots$ is a sequence such that $x_{n}\rightarrow x$
then 
\[
\lim_{n\to\infty}\left(REG_{Q}\left(\left.x^{n}\right|x^{m}\right)-\frac{k}{2}\ln\frac{n}{\tau}\right)=\ln\left(J\left|x^{m}\right.\right).
\]

\end{thm}
Since Jeffreys prior has regret that is asymptotically constant and
since according to Theorem \ref{thm:relative} one prediction system
cannot be uniformly better than another we see that an exponential
prediction system based on Jeffreys prior is optimal with respect
to regret in the following sense.
\begin{cor}
For any exponential prediction system there exists an element $P^{x}$
in the exponential family corresponding to an interior point $x$
in the convex core and a sequence $x_{1}x_{2}\dots$ such that $x_{n}\rightarrow x$
such that the regret of the exponential prediction system satisfies
\[
\lim_{n\to\infty}\inf\left(REG\left(\left.x^{n}\right|x^{m}\right)-\frac{k}{2}\ln\frac{n}{\tau}\right)\geq\ln\left(J\left|x^{m}\right.\right).
\]

\end{cor}
This theorem has important consequences. For instance it becomes much
easier to prove the recent result that the SNML predictor is exchangable
if and only if it is equivalent to the use for Jeffreys prior \cite{Hedayati2012}.

\subsection{When is conditional Jeffreys Finite?}

After having identified Jeffreys prior as optimal it is of interest
to see how long sequences are needed before the conditional Jeffreys
integral becomes finite. Most exponential families used in applications
have finite conditional Jeffreys integral after just one sample point.
For a one dimensional exponential family one can divide the parameter
interval into a left part and a right part and treat these independently.
The following results seem to cover all cases relevant for applications.
\begin{thm}
\label{thm:Ana}Let $Q$ be a measure for which the convex core is
lower bounded. Assume that $a$ is the left end point of $M$. If
$Q$ has density $f\left(x\right)=\left(x-a\right)^{\gamma-1}g\left(x\right)$
in an interval just to the right of $a$ where $g$ is an analytic
function and $g\left(a\right)>0$ then the conditional Jeffreys integral
of the right truncated exponential family is finite.
\end{thm}
Gr{\"u}nwald and Harremo{\"e}s have previously shown that under the conditions
of the previous theorem if there is a point mass in $a$ then the
unconditional Jeffreys integral is also finite \cite{Grunwald2009}. 
\begin{thm}
\label{thm:Light}Let $(\Gamma_{0}^{\text{can}},Q)$ represent a left-truncated
exponential family that is light tailed in the sense that there exists
a Gamma exponential family such that the variance function $V$ of
$(\Gamma_{0}^{\text{can}},Q)$ satisfy 
\[
\lim\inf_{x\to\infty}\frac{V\left(x\right)}{V_{\gamma}\left(x\right)}>0
\]
then the conditional Jeffreys integral is finite where $V_{\gamma}\left(x\right)$
denotes the variance function of the gamma exponential family.
\end{thm}
The following theorem extends a theorem from \cite{Grunwald2009}.
\begin{thm}
\label{prop:betacasejeffinite} Let $(\Gamma_{0}^{\text{can}},Q)$
represent a left-truncated exponential family such that $\beta_{\sup}=0$
and $Q$ admits a density $q$ either with respect to Lebesgue measure
or counting measure. If $q$ is heavy tailed the Jeffreys integral
is finite, if and only if all the conditional Jeffreys integrals are
finite. If $q(x)=O(x^{\textrm{-}1-\alpha})$ for some $\alpha>0$,
then Jeffreys integral $\int_{M}V(x)^{\textrm{-}\nicefrac{1}{2}}\,\mathrm{d}x$
is finite. 
\end{thm}
Most exponential families with finite minimax regret also have finite
Jeffreys integral but there are counter examples and they give exponential
families for which the Jeffreys integral is always infinite.
\begin{example}
\label{Cauchy}If $Y$ is a Cauchy distributed random variable then
$X=\exp\left(Y\right)$ has a very heavy tailed distribution that
we will call a exponentiated Cauchy distribution. A probability measure
$Q$ is defined as a $\nicefrac{1}{2}$ and $\nicefrac{1}{2}$ mixture
of a point mass in 0 and an exponentiated Cauchy distribution. As
shown by Gr{\"u}nwald and Harremo{\"e}s \cite{Grunwald2009} the exponential
family based on $Q$ has finite minimax regret, but infinite Jeffreys
integral. Since the minimax regret is finite the divergence is bounded
and the conditional Jeffreys integrals are all infinite for any initial
sequence $x^{m}$ of any length.
\end{example}

\section{Discussion\label{sec:Discussion}}

The notion of sufficiency has been generalized by S. Lauritzen \cite{Lauritzen1982a}
and generalizations of his results to the setting presented here is
highly relevant but cannot be covered in this short note. In cases
where the Jeffreys integral is infinite and the minimax regret is
finite one cannot find an optimal exponential prediction system, so
exchangability cannot be achieved. In such cases the usual NML predictor
or the SNML predictor may be good alternatives. Much of what has been
said here about regret will also hold for mean redundancy \cite{Liang2004}
or for any capacity of order $\alpha$ as defined in \cite{Erven2012}.

\section*{Acknowledgement}

The author thank Peter Gr{\"u}nwald, Fares Hedayati, Wojciech Kot\l{}owski
and Peter Bartlett for stimulating discussions. Peter Gr{\"u}nwald and
Wojciech Kot\l{}owski have also provided useful comments to a previous
version of this manuscript.

\bibliographystyle{ieeetr}
\bibliography{C:/Matematik/bibtex/database1}
\pagebreak{}

\section{Appendix}

\subsection{Proof of Theorem \ref{thm:Kraft}}

Assume that $\ell$ satisfies Kraft's inequality. Then
\[
\sum_{a_{1}a_{2}...a_{n}\in\mathbb{A}^{n}}\beta^{\textrm{-}\sum_{i=1}^{n}\ell\left(a_{i}\right)}=\left(\sum_{a\in\mathbb{A}}\beta^{\textrm{-}\ell\left(a\right)}\right)^{n}\leq1^{n}=1.
\]
Therefore the function $\tilde{\ell}:\mathbb{A}^{n}\rightarrow\mathbb{N}$
given by 
\[
\tilde{\ell}\left(a_{1}a_{2}...a_{n}\right)=\left\lceil \sum_{i=1}^{n}\ell\left(a_{i}\right)\right\rceil 
\]
is integer valued and satisfies Kraft's inequality and there exists
a prefix-free code $\kappa:\mathbb{A}^{n}\rightarrow\left\{ 0,1\right\} ^{\ast}$
such that $\ell_{\kappa}\left(a_{1}a_{2}...a_{n}\right)=\tilde{\ell}\left(a_{1}a_{2}...a_{n}\right).$
Therefore
\begin{multline*}
\left\vert \bar{\ell}_{\kappa}\left(a_{1}a_{2}...a_{n}\right)-\frac{1}{n}\sum_{i=1}^{n}\ell\left(a_{i}\right)\right\vert \\
=\frac{1}{n}\left\vert \left\lceil \sum_{i=1}^{n}\ell\left(a_{i}\right)\right\rceil -\sum_{i=1}^{n}\ell\left(a_{i}\right)\right\vert \leq\frac{1}{n}
\end{multline*}
for any $\varepsilon>0$ choose $n$ such that $\nicefrac{1}{n}\leq\varepsilon.$ 

Assume that for all $\varepsilon>0$ there exists a uniquely decodable
fixed-to-variable length code $\kappa:\mathbb{A}^{n}\rightarrow\left\{ 0,1\right\} ^{\ast}$
such that 
\[
\left\vert \bar{\ell}_{\kappa}\left(a_{1}a_{2}...a_{n}\right)-\frac{1}{n}\sum_{i=1}^{n}\ell\left(a_{i}\right)\right\vert \leq\varepsilon
\]
 for all strings $a_{1}a_{2}...a_{n}\in\mathbb{A}^{n}.$ Then $n\bar{\ell}_{\kappa}\left(a_{1}a_{2}...a_{n}\right)$
satisfies Kraft's Inequality and
\begin{align*}
\left(\sum_{a\in\mathbb{A}}\beta^{\textrm{-}\ell\left(a\right)}\right)^{n} & =\sum_{a_{1}a_{2}...a_{n}\in\mathbb{A}^{n}}\beta^{\textrm{-}\sum_{i=1}^{n}\ell\left(a_{i}\right)}\\
 & \leq\sum_{a_{1}a_{2}...a_{n}\in\mathbb{A}^{n}}\beta^{\textrm{-}n\left(\bar{\ell}_{\kappa}\left(a_{1}a_{2}...a_{n}\right)-\varepsilon\right)}\\
 & =\beta^{n\varepsilon}\sum_{a_{1}a_{2}...a_{n}\in\mathbb{A}^{n}}\beta^{\textrm{-}n\bar{\ell}_{\kappa}\left(a_{1}a_{2}...a_{n}\right)}\\
 & \leq\beta^{n\varepsilon}.
\end{align*}
Therefore$\sum_{a\in\mathbb{A}}\beta^{\textrm{-}\ell\left(a\right)}\leq\beta^{\varepsilon}$
for all $\varepsilon>0$ and the result is obtained.

\subsection{Proof of Theorem \ref{thm:The-sets-convex}}

First we will prove that $F_{n}$ is convex. Assume that $x_{0},x_{1}\in F_{n}.$
Then 
\[
\intop_{cc}\exp\left(\textrm{-}nD\left(\left.x_{i}\right\Vert x\right)\right)\,\mathrm{d}\nu x<\infty.
\]
 For $s\in\left[0,1\right]$ introduce $x_{s}=\left(1-s\right)x_{0}+sx_{1}.$
Then 
\begin{align*}
D\left(\left.x_{s}\right\Vert x\right)= & \left(1-s\right)D\left(\left.x_{0}\right\Vert x\right)+sD\left(\left.x_{1}\right\Vert x\right)\\
 & -\left(\left(1-s\right)D\left(\left.x_{0}\right\Vert x_{s}\right)+sD\left(\left.x_{1}\right\Vert x_{s}\right)\right)\\
\geq & \left(1-s\right)D\left(\left.x_{0}\right\Vert x\right)+sD\left(\left.x_{1}\right\Vert x\right)\\
 & -C\left(P^{x_{0}},P^{x_{1}}\right),
\end{align*}
 where $C\left(P,Q\right)$ denotes the Chernoff information between
$P$ and $Q.$ Hence
\begin{multline*}
\intop_{M}\exp\left(\textrm{-}nD\left(\left.x_{s}\right\Vert x\right)\right)\,\mathrm{d}\nu x\\
\leq\intop_{M}\exp\left(\textrm{-}n\left(\begin{array}{c}
\left(1-s\right)D\left(\left.x_{0}\right\Vert x\right)+sD\left(\left.x_{1}\right\Vert x\right)\\
-C\left(P^{x_{0}},P^{x_{1}}\right)
\end{array}\right)\right)\,\mathrm{d}\nu x\\
\leq\mathrm{e}^{nC\left(P^{x_{0}},P^{x_{1}}\right)}\intop_{M}\exp\left(\textrm{-}n\left(\begin{array}{c}
\left(1-s\right)D\left(\left.x_{0}\right\Vert x\right)\\
+sD\left(\left.x_{1}\right\Vert x\right)
\end{array}\right)\right)\,\mathrm{d}\nu x\\
\leq\mathrm{e}^{nC\left(P^{x_{0}},P^{x_{1}}\right)}\left(\begin{array}{c}
\left(1-s\right)\intop_{M}\exp\left(-nD\left(\left.x_{0}\right\Vert x\right)\right)\,\mathrm{d\nu}x\\
+s\intop_{M}\exp\left(-nD\left(\left.x_{1}\right\Vert x\right)\right)\,\mathrm{d}\nu x
\end{array}\right)\\
<\infty.
\end{multline*}

Next we note that $\exp\left(-nD\left(\left.x_{0}\right\Vert x\right)\right)$
is decreasing in n, which proves that the sequence of sets $F_{n}$
is increasing.

\subsection{Proof of Theorem \ref{thm:Conv}}

Let $x_{s}=\left(1-\frac{m}{n}\right)x_{0}+\frac{m}{n}x_{1}$. Then
\begin{multline*}
D\left(\left.x_{s}\right\Vert x\right)=D\left(\left.P^{x_{s}}\right\Vert P^{x}\right)\\
=\left(1-\frac{m}{n}\right)D\left(\left.P^{x_{0}}\right\Vert P^{x}\right)+\frac{m}{n}D\left(\left.P^{x_{1}}\right\Vert P^{x}\right)\\
-\left(\left(1-\frac{m}{n}\right)D\left(\left.P^{x_{0}}\right\Vert P^{x_{s}}\right)+\frac{m}{n}D\left(\left.P^{x_{1}}\right\Vert P^{x_{s}}\right)\right)\\
\geq\frac{m}{n}D\left(\left.x_{1}\right\Vert x\right)-C\left(P^{x_{0}},P^{x_{1}}\right),
\end{multline*}
where $C\left(P,Q\right)$ denotes the Chernoff information between
$P$ and $Q.$ Hence 
\begin{multline*}
\intop_{M}\exp\left(\textrm{-}nD\left(\left.x_{s}\right\Vert x\right)\right)\,\mathrm{d}\nu x\leq\\
\exp\left(nC\left(P^{x_{0}},P^{x_{1}}\right)\right)\intop_{M}\exp\left(\textrm{-}mD\left(\left.x_{1}\right\Vert x\right)\right)\,\mathrm{d}\nu x<\infty.
\end{multline*}

\subsection{Proof of Theorem \ref{thm:Laur}}

Consider an exponential prediction system $P\left(\cdot\vert\cdot\right)$
with sufficient statistic $X$ with values in $\mathbb{R}^{d}.$ First
we will assume that $P\left(\cdot\vert\cdot\right)$ is defined for
any initializing sequence of length zero. Therefore we consider a
probability measure $P$ on finite sequences $X_{1},X_{2},\dots,X_{n}$
such that the distribution of $X_{n+1}$ is independent of $X_{1}^{n}$
given the value of $S_{n}=\frac{1}{n}\left(X_{1}+X_{2}+\dots+X_{n}\right).$
Then the distribution of $X_{1}^{n}$ is independent of $S_{n+1}$
given $S_{n}$. Let $\mathit{\mathcal{M}}$ denote the set of all
probability measures on sequences such that conditional distribution
of $X^{n}$ given $S_{n}$ equals the conditional distrubion generated
by $P.$ S. Lauritzen called the set $\mathcal{M}$ a \emph{maximal
family} \cite{Lauritzen1982a} and he proved that this is a Choquet
simplex. Let $\mathcal{E}$ denote the expreme points of this simplex.
Our goal is to identfy these extreme points.

Let $Q$ denote a distribution in the maximal family. Then the sequence
$S_{n}$ is a reverse martingale in the sense that each coordinate
of the random vector is a reversed martingale. We know that a reversed
martingale converges almost surely to a random variable $S_{\infty}$
on the tail algebra generated by $S_{1},S_{2},\dots$ Therefore the
distribution $Q$ can be decomposed as a mixture of distribution each
corresponding to a possible value of $S_{\infty}.$ We have that $E\left[S_{n}\right]=E\left[S_{\infty}\right]$
for all $n$ so if the measure on $S_{\infty}$ is concentrated in
a point then this point equals $E\left[S_{1}\right]=E\left[X_{1}\right]$
which is an interior point in the convex core of $Q$ restricted to
$X_{1}.$

Next we shall prove that $X_{1},X_{2},\dots,X_{n}$ are independent
given $S_{\infty}.$ It is sufficient to prove that $X^{n-1}$ is
independent of $X_{n}$ given $S_{\infty}.$ We have that $X^{n-1}$
is independent of $X_{n}$ given $\frac{1}{\ell}\left(X_{1}+X_{2}+\dots+X_{n-1}+X_{n+1}+\dots X_{\ell+1}\right)$
but this random variable converges to $S_{\infty}$ for $\ell$ tending
to $\infty$ and the result follows.

Let $Q_{1}$ and $Q_{2}$ denote two extreme elements of the maximal
family. Then $Q_{1}$ and $Q_{2}$ have the same restriction to $X_{1}^{n}$
given $S_{n}$ which implies that 
\[
\frac{\mathrm{d}Q_{1}}{\mathrm{d}Q_{2}}\left(X^{n}\right)
\]
only depends on the value of $S_{n}.$ Since $X_{1},X_{2},\dots,X_{n}$
are independent under $Q_{1}$ and under $Q_{2}$ we have that
\[
\ln\left(\frac{\mathrm{d}Q_{1}}{\mathrm{d}Q_{2}}\left(X^{n}\right)\right)=\sum_{i=1}^{n}\ln\left(\frac{\mathrm{d}Q_{1}}{\mathrm{d}Q_{2}}\left(X_{i}\right)\right)
\]
 which implies that $\ln\left(\frac{\mathrm{d}Q_{1}}{\mathrm{d}Q_{2}}\left(x\right)\right)$
is a linear function of $x$. Hence $Q_{1}$ and $Q_{2}$ are two
elements of an exponential family with $x$ as sufficient statistic.
We also see that $S_{\infty}$ may be identified with the mean value
of the distribution $Q$ restricted to $X_{1}.$ Hence the predictor
$P$ is a mixture of elements in an exponential family.

In general we should take the conditioning sequence into account.
For any initial sequence $x_{1}^{m}$ we get a distribution $\eta_{x_{1}^{m}}$
over the mean value range of an exponential family. Let $y_{1}^{\ell}$
denote another initial sequence. Then $\eta_{x_{1}^{m}}$ conditioned
on $y_{1}^{\ell}$ will equal $\eta_{y_{1}^{\ell}}$ conditioned on
$x_{1}^{m}$. Hence 
\[
\frac{\mathrm{d}\eta_{x_{1}^{m}}}{\mathrm{d}\eta_{y_{1}^{\ell}}}\left(z\right)\cdot\frac{\frac{\mathrm{e}^{\textrm{\textrm{-}}\ell D\left(\left.\bar{y}\right\Vert z\right)}}{\int_{M}\mathrm{e}^{\textrm{-\ensuremath{\ell}}D\left(\left.\bar{y}\right\Vert z\right)}\,\mathrm{d}\eta_{x_{1}^{m}}z}}{\frac{\mathrm{e}^{\textrm{-}nD\left(\left.\bar{x}\right\Vert z\right)}}{\int_{M}\mathrm{e}^{\textrm{\textrm{-}}nD\left(\left.\bar{x}\right\Vert z\right)}\,\mathrm{d}\eta_{y_{1}^{\ell}}z}}=1.
\]
From this we see that
\begin{multline*}
\frac{\mathrm{d}\eta_{x_{1}^{m}}}{\mathrm{d}\eta_{y_{1}^{\ell}}}\left(z\right)\cdot\frac{\mathrm{e}^{\textrm{-}nD\left(\left.\bar{x}\right\Vert z\right)}}{\mathrm{e}^{\textrm{-}\ell D\left(\left.\bar{y}\right\Vert z\right)}}=\frac{\int_{M}\mathrm{e}^{\textrm{-}\ell D\left(\left.\bar{y}\right\Vert z\right)}\,\mathrm{d}\eta_{x_{1}^{m}}z}{\int_{M}\mathrm{e}^{\textrm{-n}D\left(\left.\bar{x}\right\Vert z\right)}\,\mathrm{d}\eta_{y_{1}^{\ell}}z}
\end{multline*}
and therefore the measures 
\[
\eta_{x_{1}^{m}}\cdot\mathrm{e}^{\textrm{\textrm{-}}nD\left(\left.\bar{x}\right\Vert z\right)}\sim d\eta_{y_{1}^{\ell}}\cdot\mathrm{e}^{\textrm{-}\ell D\left(\left.\bar{y}\right\Vert z\right)}.
\]
That means that the measure $\eta_{x_{1}^{m}}\cdot\exp\left(D\left(\left.\bar{x}\right\Vert z\right)\right)$
does not depend on the initializing sequence except for a constant
factor. Let $\eta$ denote one of these measure that may or may not
be normalized. We have that
\[
\eta_{x_{1}^{m}}\cdot\mathrm{e}^{\textrm{-}nD\left(\left.\bar{x}\right\Vert z\right)}\sim\eta
\]
and therefore
\[
\frac{\mathrm{d}\eta_{x_{1}^{m}}}{\mathrm{d}\eta}\left(z\right)=\frac{\mathrm{e}^{\textrm{\textrm{-}}nD\left(\left.\bar{x}\right\Vert z\right)}}{\int_{M}\mathrm{e}^{\textrm{\textrm{-}}nD\left(\left.\bar{x}\right\Vert z\right)}\,\mathrm{d}\eta}
\]
for any initial sequence $x_{1}^{m}$ for which $P\left(\cdot\vert x_{1}^{m}\right)$
is defined. Finally we get
\begin{multline*}
\frac{\mathrm{d}P\left(\cdot\left|x^{m}\right.\right)}{\mathrm{d}P_{0}}\left(x\right)=\\
\int_{M}\frac{\exp\left(x\cdot\hat{\beta}\left(z\right)\right)}{Z\left(\hat{\beta}\left(z\right)\right)}\,\mathrm{d}\eta_{x_{1}^{m}}z=\\
\int_{M}\frac{\exp\left(x\cdot\hat{\beta}\left(z\right)\right)}{Z\left(\hat{\beta}\left(z\right)\right)}\frac{\mathrm{e}^{\textrm{-}nD\left(\left.\bar{x}\right\Vert z\right)}}{\int_{M}\mathrm{e}^{\textrm{-n}D\left(\left.\bar{x}\right\Vert z\right)}\,\mathrm{d}\eta z}\,\mathrm{d}\eta z\,.
\end{multline*}

\subsection{Proof of Theorem \ref{thm:relative}}

Assume that the exponential prediction systems $Q_{1}$ and $Q_{2}$
are based on priors $\mu$ and $\nu$. Find initial data sequences
for $\mu$ and for $\nu$ that allow the prior measures to be normalized.
A concatenation of these two initial sequences into a sequence $x^{m}$
that allow both $\mu$ and $\nu$ to be normalized. Without loss of
generality we will assume that the initializing sequence has length
zero. Therefore we will assume that $\mu$ and $\nu$ are probability
measures.

Let $Q_{1}$ and $Q_{2}$ denote two different exponential prediction
systems for the same exponential family. Then 
\begin{equation}
\frac{\mathrm{d}Q_{1}}{\mathrm{d}Q_{2}}\left(x^{n}\right)=\frac{\int\exp\left(\textrm{\textrm{-}}nD\left(\left.\bar{x}\right\Vert z\right)\right)\,\textrm{d}\mu z}{\int\exp\left(\textrm{\textrm{-}}nD\left(\left.\bar{x}\right\Vert z\right)\right)\,\textrm{d}\nu z}.\label{eq:kvotient}
\end{equation}
Let $\tilde{\mu}$ denotes the absolutely continuous part of $\mu$
with respect to $\nu.$ Then 
\[
\frac{\int\exp\left(\textrm{\textrm{-}}nD\left(\left.\bar{x}\right\Vert z\right)\right)\,\textrm{d}\mu z}{\int\exp\left(\textrm{\textrm{-}}nD\left(\left.\bar{x}\right\Vert z\right)\right)\,\textrm{d}\nu z}\to\frac{\textrm{d}\tilde{\mu}}{\textrm{d}\nu}\left(\bar{x}\right).
\]
$\nu$-almost surely for $n$ tending to $\infty.$ Since
\[
\int\frac{\textrm{d}\tilde{\mu}}{\textrm{d}\nu}\left(z\right)\,\textrm{d}\nu z\leq1
\]
and $\mu\neq\nu$ there exists a sequence $x_{1},x_{2},\dots$ converging
to some $z$ such that $\lim_{n\to\infty}\frac{dP}{dQ}\left(x^{n}\right)<1.$
Hence the regret of $Q_{2}$ is greater than the regret of $Q_{1}$
for this sequence.

\subsection{Proof of Theorem \ref{thm:taet}}

Let $x_{1},x_{2},\dots$ denote a sequence satisfying 
\[
\lim\inf D\left(\left.P^{\bar{x}}\right\Vert P^{x}\right)>0.
\]
Assume without loss of generality that 
\[
D\left(\left.P^{\bar{x}}\right\Vert P^{x}\right)\geq\delta>0
\]
for all $n.$ Assume that $Q_{0}=P_{0}$ and that the support of the
prior measure $\nu$ equals the closure of the mean value range of
the exponential family. First assume that the conditioning sequence
has length zero. Then the regret of $x^{m}$ is 
\begin{multline*}
\frac{1}{n}\ln\frac{\int\exp\left(\textrm{-}nD\left(\left.P^{\bar{x}}\right\Vert P^{y}\right)\right)\,\textrm{d}\nu y}{\exp\left(\textrm{-}nD\left(\left.P^{\bar{x}}\right\Vert P^{x}\right)\right)}\\
\geq\frac{\delta}{2}+\frac{1}{n}\ln\left(\nu\left\{ y\left|D\left(\left.P^{\bar{x}}\right\Vert P^{y}\right)<D\left(\left.P^{\bar{x}}\right\Vert P^{x}\right)-\frac{\delta}{2}\right.\right\} \right)\\
=\frac{1}{n}\ln\left(\int\exp\left(-n\left(D\left(\left.P^{\bar{x}}\right\Vert P^{x}\right)-D\left(\left.P^{\bar{x}}\right\Vert P^{y}\right)\right)\right)\,\textrm{d}\nu y\right)\\
\geq\frac{1}{n}\ln\left(\int_{D_{n}}\exp\left(-n\left(D\left(\left.P^{\bar{x}}\right\Vert P^{x}\right)-D\left(\left.P^{\bar{x}}\right\Vert P^{y}\right)\right)\right)\,\textrm{d}\nu y\right)
\end{multline*}
where $D_{n}$ denotes the set
\[
\left\{ y\left|D\left(\left.P^{\bar{x}}\right\Vert P^{y}\right)<D\left(\left.P^{\bar{x}}\right\Vert P^{x}\right)-\frac{\delta}{2}\right.\right\} .
\]

The set $D_{n}$ decreases as $\bar{x}$ gets closer to $x$ so we
may without loss of generality assume that $D\left(\left.P^{\bar{x}}\right\Vert P^{x}\right)=\delta$.
Now we just have to remark that
\begin{multline*}
\nu\left\{ y\left|D\left(\left.P^{\bar{x}}\right\Vert P^{y}\right)<D\left(\left.P^{\bar{x}}\right\Vert P^{x}\right)-\frac{\delta}{2}\right.\right\} \\
=\nu\left\{ y\left|D\left(\left.P^{\bar{x}}\right\Vert P^{y}\right)<\frac{\delta}{2}\right.\right\} 
\end{multline*}
 is positive for all values of $\bar{x}$ and has a minimum because
$\nu\left\{ y\left|D\left(\left.P^{\bar{x}}\right\Vert P^{y}\right)<\frac{\delta}{2}\right.\right\} $
is a continuous function of $\bar{x}$ over the compact set $\left\{ \bar{x}\left|D\left(\left.P^{\bar{x}}\right\Vert P^{x}\right)=\delta\right.\right\} $. 

The conditional version of the theorem follows because a prior measure
and a posterior measure are mutually absolutely continuous.

In the exponential family corresponding to $Q$ there exists a distribution
that is closest to $P_{0}$. We will denote this distribution $Q_{0}.$
If $Q_{0}=P_{0}$ then the two exponential families are equal. If
$Q_{0}\neq P_{0}$ then 
\[
D\left(P_{0}\left\Vert Q_{0}\right.\right)>0.
\]
 Let $P^{x}$ denote an element in the exponential family such that
$D\left(P^{x}\left\Vert P_{0}\right.\right)<D\left(P^{x}\left\Vert Q^{x}\right.\right).$
Then the sequence $x,x,x,\dots$has regret bounded by
\begin{multline*}
\frac{1}{n}\ln\frac{\int\exp\left(\textrm{-}nD\left(\left.P^{\bar{x}}\right\Vert Q^{y}\right)\right)\,\textrm{d}\nu y}{\exp\left(\textrm{-}nD\left(\left.P^{\bar{x}}\right\Vert P^{\mu_{0}}\right)\right)}\\
=\frac{1}{n}\ln\frac{\int\exp\left(\textrm{-}nD\left(\left.P^{x}\right\Vert Q^{y}\right)\right)\,\textrm{d}\nu y}{\exp\left(\textrm{-}nD\left(\left.P^{x}\right\Vert P^{\mu_{0}}\right)\right)}\\
\leq\frac{1}{n}\ln\frac{\int\exp\left(\textrm{-}nD\left(\left.P^{x}\right\Vert Q^{x}\right)\right)\,\textrm{d}\nu y}{\exp\left(\textrm{-}nD\left(\left.P^{x}\right\Vert P_{0}\right)\right)}\\
=D\left(\left.P^{x}\right\Vert P_{0}\right)-D\left(\left.P^{x}\right\Vert Q^{x}\right),
\end{multline*}
so coding by $P_{0}$ is better than coding by the exponential predition
system $Q$ by a certain constant. Assume that $P_{0}=Q_{0}$. Then
the two esponential families are equal. Assume that $\nu$ is not
dense. Let $x$ denote an element in the mean value range $M$ such
that $\nu\left\{ y\mid D\left(P^{x}\Vert P^{y}\right)<r\right\} =0$
for some $r>0.$ Let $P^{z}$ denote an element in the exponential
family such that $D\left(P^{x}\Vert P^{z}\right)<r$. Then the sequence
$x,x,x,\dots$has regret bounded by
\begin{multline*}
\frac{1}{n}\ln\frac{\int\exp\left(\textrm{-}nD\left(\left.P^{x}\right\Vert Q^{y}\right)\right)\,\textrm{d}\nu y}{\exp\left(\textrm{-}nD\left(\left.P^{x}\right\Vert P^{x}\right)\right)}\\
\leq\frac{1}{n}\ln\frac{\int\exp\left(-nr\right)\,\textrm{d}\nu y}{\exp\left(\textrm{-}nD\left(\left.P^{x}\right\Vert P^{x}\right)\right)}\\
=D\left(\left.P^{x}\right\Vert P^{x}\right)-r<0.
\end{multline*}

\subsection{Proof of Theorem \ref{thm:Ana}}
\begin{lem}
For an exponential family the natural parameter $\beta$, the cumulant
generating function $A\left(\beta\right)$, and the divergence can
be calculated from the variance function $V$ as follows; where the
variance function is a mapping from the mean of the family to its
variance. 
\begin{align}
\hat{\beta}\left(\mu\right) & =\int\frac{1}{V\left(\mu\right)}\,\mathrm{d}\mu,\label{eq:NatLig}\\
A\left(\hat{\beta}\left(\mu\right)\right) & =\int\frac{\mu}{V\left(\mu\right)}\,\mathrm{d}\mu,\label{eq:KumLig}\\
D\left(\left.\mu_{0}\right\Vert \mu_{1}\right) & =\int_{\mu_{0}}^{\mu_{1}}\frac{\mu-\mu_{0}}{V\left(\mu\right)}\,\mathrm{d}\mu.\label{eq:DivLig}
\end{align}
\end{lem}
\begin{IEEEproof}
We use that $A\left(\beta\right)$ is the cumulant generating function,
so that $\frac{\mathrm{d}A\left(\beta\right)}{\mathrm{d}\beta}=\mu\left(\beta\right)$
and $\frac{\mathrm{d}^{2}A\left(\beta\right)}{\mathrm{d}\beta^{2}}=V\left(\mu\left(\beta\right)\right).$
Hence $\frac{\mathrm{d}\mu}{\mathrm{d}\beta}=V\left(\mu\left(\beta\right)\right)$
from which the first Equation \ref{eq:NatLig} follows. 

We have 
\[
\frac{\mathrm{d}A\left(\hat{\beta}\left(\mu\right)\right)}{\mathrm{d}\mu}=\frac{\frac{\mathrm{d}A\left(\theta\right)}{\mathrm{d}\beta}}{\frac{\mathrm{d}\mu}{\mathrm{d}\beta}}=\frac{\mu}{V\left(\mu\right)}
\]
 from which Equation \ref{eq:KumLig} follows.

The divergence is given by 
\begin{multline}
D\left(\left.P_{\beta_{0}}\right\Vert P_{\beta_{1}}\right)=E\left[\ln\left(\frac{\mathrm{d}P_{\beta_{0}}}{\mathrm{d}P_{\beta_{1}}}\right)\right]=\\
E_{P_{\theta_{0}}}\left[\ln\left(\frac{\exp\left(\beta_{0}\cdot X-A\left(\beta_{0}\right)\right)}{\exp\left(\beta_{1}\cdot X-A\left(\beta_{1}\right)\right)}\right)\right]\\
=\left(\beta_{0}\cdot\mu\left(\beta_{0}\right)-A\left(\beta_{0}\right)\right)-\left(\beta_{1}\cdot\mu\left(\beta_{0}\right)-A\left(\beta_{1}\right)\right).
\end{multline}
 The derivative with respect to $\beta_{1}$ is 
\[
\frac{\mathrm{d}}{\mathrm{d}\beta_{1}}D\left(\left.P_{\beta_{0}}\right\Vert P_{\beta_{1}}\right)=\mu\left(\beta_{1}\right)-\mu\left(\beta_{0}\right)
\]
 Hence the derivative with respect to $\mu_{1}=\mu\left(\theta_{1}\right)$
is 
\[
\frac{\mathrm{d}}{\mathrm{d}\mu_{1}}D\left(\left.P_{\beta_{0}}\right\Vert P_{\beta_{1}}\right)=\frac{\mu\left(\beta_{1}\right)-\mu\left(\beta_{0}\right)}{V\left(\mu\left(\beta_{1}\right)\right)}.
\]
 Together with the obvious fact that 
\[
D\left(\left.\mu_{0}\right\Vert \mu_{0}\right)=\int_{\mu_{0}}^{\mu_{0}}\frac{\mu-\mu_{0}}{V\left(\mu\right)}\,\mathrm{d}\mu
\]
 Equation \ref{eq:DivLig} follows.
\end{IEEEproof}
The variance function can be approximated by $V\left(\mu\right)\approx c_{0}\left(\mu-\mu_{\inf}\right)^{p},$
where $p=2$ if there is no point mass in $\mu_{\inf}$ and $p<2$
if there is a point mass in $\mu_{\inf}$ \cite[Thm. 4.4]{jorgensen1997}.
Therefore the integrand in the Jeffreys integral can be approximated
by $c_{0}^{-\nicefrac{p}{2}}\left(x-\mu_{\inf}\right)^{-\nicefrac{p}{2}}$
near $\mu_{\inf}$ so the left endpoint gives a finite contribution
to the Jeffreys integral if and only if $p<2$. 

Assume that there is no point mass in $\mu_{\inf}$ and that $\mu_{\inf}=0,$
so that the integrand in the Jeffreys integral can be approximated
by $c_{0}^{-\nicefrac{p}{2}}x^{-1}$ near 0. According to \ref{eq:DivLig}
the divergence can be calculated from the variance function as
\begin{align*}
D\left(\left.\mu_{1}\right\Vert \mu_{2}\right) & =\int_{\mu_{1}}^{\mu_{2}}\frac{\mu-\mu_{1}}{V\left(\mu\right)}\,\mathrm{d}\mu\\
 & \approx\int_{\mu_{2}}^{\mu_{1}}\frac{\mu_{1}-\mu}{c_{0}\mu^{2}}\,\mathrm{d}\mu\\
 & =c_{0}^{-1}\left[-\frac{\mu_{1}}{\mu}-\ln\left(\mu\right)\right]_{\mu_{2}}^{\mu_{1}}\\
 & =c_{0}^{-1}\left(\frac{\mu_{1}}{\mu_{2}}-1-\ln\left(\frac{\mu_{1}}{\mu_{2}}\right)\right).
\end{align*}
Hence the conditional Jeffreys integral is 
\begin{multline*}
\int_{0}^{\mu_{3}}c_{0}^{-\nicefrac{1}{2}}x^{-1}\exp\left(-nc_{0}^{-1}\left(\frac{\mu_{1}}{x}-1-\ln\left(\frac{\mu_{1}}{x}\right)\right)\right)\,\mathrm{d}x =\\
\mu_{1}^{^{nc_{0}^{-1}}}c_{0}^{-\nicefrac{1}{2}}\exp\left(nc_{0}^{-1}\right)\int_{0}^{\mu_{3}}x^{-1-nc_{0}^{-1}}\exp\left(-\frac{nc_{0}^{-1}\mu_{1}}{x}\right)\, dx.
\end{multline*}
In the last integral we make the substitution $t=x^{-1}$ leading
to
\begin{multline*}
\int_{0}^{\mu_{3}}x^{-1-nc_{0}^{-1}}\exp\left(-\frac{nc_{0}^{-1}\mu_{1}}{x}\right)\, dx \\ =\int_{\mu_{3}^{-1}}^{\infty}t^{nc_{0}^{-1}-1}\exp\left(-nc_{0}^{-1}\mu_{1}t\right)\,\mathrm{d}x <\infty.
\end{multline*}

\subsection{Proof of Theorem \ref{thm:Light}}

The Gamma exponential families have finite conditional Jeffreys integral.
We use the Equation \ref{eq:DivLig} and the formula for the conditional
Jeffreys integral 
\[
\int_{M}\frac{\exp\left(\textrm{-}nD\left(\left.P^{y}\right\Vert P^{x}\right)\right)}{V\left(x\right)^{\nicefrac{1}{2}}}\,\mathrm{d}x
\]
 to conclude that a larger variance function leads to a smaller Jeffreys
integral.

\subsection{Proof of Theorem \ref{prop:betacasejeffinite}}

Assume that q is heavy tailed. Gr{\"u}nwald and Harremo{\"e}s have shown that
\cite{Grunwald2009} in this case 
\[
\sup_{y>x}D\left(\left.Q^{x}\right\Vert Q^{y}\right)<\infty,
\]
 which implies that the factor $\exp\left(\textrm{\textrm{-}}mD\left(\left.Q^{x}\right\Vert Q^{y}\right)\right)$
in the integrand of the conditional Jeffreys integral is lower bounded.
The proof of the second half of the theorem follows directly from
\cite{Grunwald2009}.

\subsection{Further detail about Example \ref{Cauchy}}

If $Y$ is a Cauchy distributed random variable then $X=\exp\left(Y\right)$
has density 
\[
\frac{2}{\tau x\left(1+\log^{2}\left(x\right)\right)}.
\]
A probability measure $Q$ is defined as a $\nicefrac{1}{2}$ and
$\nicefrac{1}{2}$ mixture of a point mass in 0 and an exponentiated
Cauchy distribution. We consider the exponential family based on $Q$.
The partition function is 
\[
Z\left(\beta\right)=\frac{1}{2}+\frac{1}{\tau}\int_{0}^{\infty}\frac{\exp\left(\beta x\right)}{x\left(1+\log^{2}\left(x\right)\right)}\,\mathrm{d}x,\ \beta\leq0.
\]
We note that $\nicefrac{1}{2}\leq Z\left(\beta\right)\leq1$ for all
$\beta\leq0.$ Then 
\[
D\left(Q_{\beta}\Vert Q\right)\leq D\left(Q_{-\infty}\Vert Q\right)=1~\text{\textrm{bit.}}
\]
Therefore the minimax redundancy is at most 1 bit.

The mean value $\mu$ as a function of $\beta$ is
\[
\mu\leq2Z^{\prime}\left(\beta\right)=\frac{2}{\tau}\int_{0}^{\infty}\frac{\exp\left(\beta x\right)}{\left(1+\log^{2}\left(x\right)\right)}\,\mathrm{d}x\leq\frac{1}{3\left\vert \beta\right\vert }.
\]
The variance as a function of $\beta$ can be lower bounded as follows:
\begin{align*}
I_{\beta} & =\frac{\frac{1}{2}\mu^{2}+\frac{1}{\tau}\int_{0}^{\infty}\left(x-\mu\right)^{2}\frac{\exp\left(\beta x\right)}{x\left(1+\log^{2}\left(x\right)\right)}\,\mathrm{d}x}{Z\left(\beta\right)}\\
 & \geq\frac{1}{\tau}\int_{\frac{2}{3}\left\vert \beta\right\vert ^{-1}}^{\left\vert \beta\right\vert ^{-1}}\left(x-\frac{1}{3\left\vert \beta\right\vert }\right)^{2}\frac{\exp\left(\beta x\right)}{x\left(1+\log^{2}\left(x\right)\right)}\,\mathrm{d}x\\
 & \geq\frac{1}{81\tau\mathrm{e}}\frac{1}{\beta^{2}\left(1+\log^{2}\left\vert \beta\right\vert \right)}.
\end{align*}
Therefore there exists a constant $c>0$ such that 

\[I_{\beta}^{\nicefrac{1}{2}}\geq c\cdot\frac{1}{\left\vert \beta\right\vert \left(1+\log^{2}\left\vert \beta\right\vert \right)^{\nicefrac{1}{2}}}.
\] 

Now assume that we have observed a sequence of length $n$ with average
$\bar{x}$. Then the posterior density is proportional to 
\[
\exp\left(-nD\left(\left.P^{\bar{x}}\right\Vert P_{\beta}\right)\right)
\]
If $\bar{x}\leq\mu_{\beta}$ then
\[
D\left(\left.P^{\bar{x}}\right\Vert P_{\beta}\right)\leq D\left(Q_{-\infty}\Vert Q\right)=\ln\left(2\right)
\]
Hence there exists a constant $\tilde{c}$ so that the posterior density
of the parameter $\beta\geq\hat{\beta}\left(\bar{x}\right)$ is lower
bounded by
\[
\tilde{c}\cdot\frac{1}{\left\vert \beta\right\vert \left(1+\log^{2}\left\vert \beta\right\vert \right)^{\nicefrac{1}{2}}}
\]
so the Jeffreys integral is infinite with an infinite contribution
from small values of $\left\vert \beta\right\vert .$ Hence, both
the left- and the right-truncated exponential family have finite minimax
regret but infinite conditional Jeffreys integral. 
\end{document}